\documentclass[12pt]{article}
\usepackage{ijmpe1,epsfig,rotating}
\makeatletter
\def\maketitle{\par
 \begingroup
 \def\thefootnote{\fnsymbol{footnote}}
 \def\@makefnmark{\mbox{$^\@thefnmark$}}
 \@maketitle 
 \@thanks
 \endgroup
 \setcounter{footnote}{0}
 \let\maketitle\relax
 \let\@maketitle\relax
 \gdef\@thanks{}\gdef\@author{}\gdef\@title{}\let\thanks\relax}
\def\@maketitle{\vspace*{0.2cm} %\vbox to 1.50in
{\hsize\textwidth
 \linewidth\hsize \centering
 {\large \bf \@title \par} \vskip 0.3cm {\normalsize  \@author \par}}}
\def\thefootnote{\mbox{\noindent$\fnsymbol{footnote}$}}
\long\def\@makefntext#1{\noindent$^{\@thefnmark}$#1}
\def\ps@plain{\let\@mkboth\@gobbletwo
     \let\@oddhead\@empty
        \def\@oddfoot{\lower2pc\hbox{\fbox{\hbox to \hsize{\reset@font\rightnote\hfill {\bf \thepage}}}}}
\let\@evenhead\@empty\let\@evenfoot\@oddfoot}
\def\rightnote{{\it \jobname : submitted to  {\bf IJMPE} on \today}}%

\makeatother

\begin{document}
\copyrightheading{}                     %{Vol. 0, No. 0 (1993) 000--000}
\pub{first draft: November 10, 1999}

%\begin{flushright}
%TRI--PP--99--28\\
%Sept 1999
%\end{flushright}

\title{Constraints on a Parity-Conserving/Time-Reversal-Non-Conserving
 Interaction\footnote{Work supported in part by the Natural Sciences and Engineering Research
   Council of Canada.}}
\author{Willem T.H. van Oers\\
Thomas Jefferson National Accelerator Facility, Newport News, \\
VA  23606\\
   and\\
   Department of Physics, University of Manitoba, Winnipeg, MB, Canada 
   R3T 2N2\\
                 and\\
 TRIUMF, 4004 Wesbrook Mall, Vancouver, BC Canada V6T 2A3}
\maketitle

\begin{abstract}     
  Time-Reversal-Invariance non-conservation has for the first time been
  unequivocally demonstrated in a direct measurement at CPLEAR. One then can
  ask the question: What about tests of time-reversal-invariance in systems
  other than the kaon system? Tests of time-reversal-invariance can be
  distinguished as belonging to two classes: the first one deals with parity
  violating (P-odd)/time-reversal-invariance-odd (T-odd) interactions, while
  the second one deals with P-even/T-odd interactions (assuming CPT
  conservation this implies C-conjugation non-conservation). Limits on a
  P-odd/T-odd interaction follow from measurements of the electric dipole
  moment of the neutron (with a present upper limit of $8 \times 10^{-26}$ e.cm
  [95\% C.L.]). It provides a limit on a P-odd/T-odd pion-nucleon coupling
  constant which is less than $10^{-4}$ times the weak interaction strength.
  Experimental limits on a P-even/T-odd interaction are much less stringent.
  Following the standard approach of describing the nucleon-nucleon
  interaction in terms of meson exchanges, it can be shown that only charged
  rho-meson exchange and A$_{1}$-meson exchange can lead to a P-even/T-odd
  interaction. The better constraints stem from measurements of the electric
  dipole moment of the neutron and from measurements of charge-symmetry
  breaking in neutron-proton elastic scattering. The latter experiments were
  executed at TRIUMF (497 and 347 MeV) and at IUCF (183 MeV). Weak decay
  experiments may provide limits which will possibly be comparable. All other
  experiments, like gamma decay experiments, detailed balance experiments,
  polarization - analyzing power difference determinations, and five-fold
  correlation experiments with polarized incident nucleons and aligned
  nuclear targets, have been shown to be at least an order of magnitude
  less sensitive. The question then emerges: is there room for further
  experimentation? 
\end{abstract}

%\begin{center}
%(paper submitted to International Journal of Modern Physics E)
%\end{center}
%\setlength{\baselineskip}{5ex}

\section{Introduction}

    Time-Reversal-Invariance non-conservation has for the first time
  \linebreak[4]been
  un\-equiv\-ocally dem\-on\-strated in a direct measure\-ment at \linebreak[4] CPLEAR.[1]
  The exper\-iment measured directly the difference in the transition
  prob\-abilities $P(\overline{K}^{0} \to K^{0})$ and $P(K^{0} \to
  \overline{K}^{0})$. A non-zero value of this difference gives model
  independent evidence of time-reversal-invariance non-conservation. The
  deduction does not depend on the validity of the $\Delta{S} = \Delta{Q}$
  rule. The result obtained for A$_T$ with

\begin{eqnarray*} 
A_T &=& \frac{R(\overline{K}^{0} \to K^{0}) - R(K^{0} \to 
\overline{K}^{0})}{R(\overline{K}^{0}
\to K^{0}) + R(K^{0} \to \overline{K}^{0})} \nonumber \\
&=& 
(6.6 \pm 1.3 {\rm (stat)} \pm 1.0 {\rm (syst)})\times 10^{-3}
\end{eqnarray*}
  is in good agreement with the measure of CP violation in neutral kaon decay.
  Starting with CPT conservation and the well established non-conservation
  of CP in kaon decays, time-reversal-invariance should also be broken. The
  CPLEAR measurement is the first direct confirmation of that. The question that
  one then can ask is: what about time-reversal-invariance non-conservation in
  systems other than the kaon system? 
 
    Tests of time-reversal-invariance can be distinguished as belonging to two
  classes: the first one deals with time-reversal-invariance-odd
  (T-odd)/parity 
violating (P-odd) interactions, while the second one deals with
  T-odd/P-even interactions (assuming CPT invariance this implies
  C-conjugation non-conservation). However, it should be noted that constraints
  on these two classes of interactions are not independent since the effects
  due to T-odd/P-odd interactions may also be produced by T-odd/P-even
  interactions in conjunction with Standard Model parity violating radiative
  corrections. The latter can occur at the $10^{-7}$ level and consequently this
  presents a limit on the constraint of T-odd/P-even interactions, which can
  be derived from experiments. Limits on a T-odd/P-odd interaction follow
  from measurements of the electric dipole moment of the neutron (which
  currently stands at $<6 \times 10^{-26}$ e.cm [95\% C.L.]). It provides a limit
  on a T-odd/P-odd pion-nucleon coupling constant which is less than
  $10^{-4}$ times the weak interaction strength.  Measurements of atomic electric
  dipole moments of $^{129}$Xe and $^{199}$Hg ( $< 8 \times 10^{-28}$ e.cm [95\% C.L.] ) give
  similar constraints. [see Ref. 2] 

    Experimental limits on a T-odd/P-even interaction are much less stringent.
  Following the standard approach of describing the nucleon-nucleon interaction
  in terms of meson exchanges, it can be shown that only charged rho-meson
  exchange and A$_{1}$-meson exchange can lead to a T-odd / P-even interaction.[3]
  The better constraints stem from measurements of the electric dipole moment of
  the neutron and from measurements of charge symmetry breaking in 
neutron-proton
  elastic scattering. All other experiments, like gamma decay experiments [4],
  detailed balance experiments [5], polarization - analyzing power difference
  measurements, and five-fold correlation experiments with polarized incident
  nucleons and aligned nuclear targets, have been shown to be at least an order
  of magnitude less sensitive. Haxton, Hoering, and Musolf [2] have deduced
  constraints on a T-odd/P-even interaction from nucleon, nuclear, and atomic
  electric dipole moments with the better constraint coming from the electric
  dipole moment of the neutron. In terms of a ratio to the strong rho-meson
  nucleon coupling constant, they deduced for the T-odd / P-even rho-meson
  nucleon coupling: $|\overline{g}_{\rho}| < 0.53 \times 10^{-3} \times 
  |f^{DDH}_\pi/f^{\rm meas.}_\pi|$.
  Here one should note that the ratio of the theoretical to the measured value
  of $f_\pi$ may be as large as 15! [6] 

    In the Standard Model a T-odd/P-even nucleon-nucleon interaction can
  hardly be accommodated. It requires C-conjugation non-conservation, which
  cannot be introduced at the first generation quark level. It can neither be
  introduced into the gluon self-interaction. Consequently one needs to consider
  C-conjugation non-conservation between quarks of different generations
  and/or between interacting fields.[7] 
   
\section{Nucleon-Nucleon Interaction}
 
    The nucleon-nucleon scattering matrix, assuming conservation of angular
  momentum, parity, time-reversal-invariance, and isospin, consists of five
  complex amplitudes, a, b, c, d, and e, which are functions of centre-of-mass
  energy E and scattering angle $\theta$. If isospin is broken (which leads to
  charge symmetry breaking in the neutron-proton system), the neutron-proton
  scattering matrix contains a sixth complex amplitude, f. If in addition one
  no longer assumes time-reversal-invariance, the neutron-proton scattering
  matrix has two additional amplitudes, g and h; the second one of these, h,
  is simultaneously time-reversal-invariance violating and charge symmetry
  breaking. [8]
\begin{eqnarray*}
 M &=& {1 \over 2}[(a + b) +
  (a - b)(\vec{\sigma}_1.\hat{n})(\vec{\sigma}_2.\hat{n}) +
  (c + d)(\vec{\sigma}_1.\hat{m})(\vec{\sigma}_2.\hat{m})  \\ \nonumber
   & & \mbox{} + (c - d)(\vec{\sigma}_1.\hat{l})(\vec{\sigma}_2.\hat{l}) +
  e((\vec{\sigma}_1.\hat{n}) + (\vec{\sigma}_2.\hat{n}))  \\ \nonumber
   & & \mbox{} + f((\vec{\sigma}_1.\hat{n}) - (\vec{\sigma}_2.\hat{n}))  
   + g((\vec{\sigma}_1.\hat{l})(\vec{\sigma}_2.\hat{m}) +
  (\vec{\sigma}_1.\hat{m})(\vec{\sigma}_2.\hat{l}))  \\ \nonumber
   & & \mbox{} +
  h((\vec{\sigma}_1.\hat{l})(\vec{\sigma}_2.\hat{m}) -
  (\vec{\sigma}_1.\hat{m})(\vec{\sigma}_2.\hat{l})), 
\end{eqnarray*}
  where $\hat{l}$, $\hat{m}$, and $\hat{n}$ are unit vectors given as
  $\hat{l}$ = $(\vec{k}_i$ + $\vec{k}_f)/|\vec{k}_i$ + $\vec{k}_f|$; $\hat{m}$ =
  $(\vec{k}_f$ - $\vec{k}_i)/|\vec{k}_f$ - $\vec{k}_i|$; $\hat{n}$ =
  $(\vec{k}_i$ $\times$ $\vec{k}_f)/|\vec{k}_i$ $\times$ $\vec{k}_f|$ 
  and $\vec{k}_i$ and $\vec{k}_f$
  are the initial and final state centre-of-mass nucleon momenta. 
  $\vec{\sigma}_1$
  and $\vec{\sigma}_2$ are Pauli spin matrices for the two nucleons.
  The proton-proton system may only contain one additional, sixth
  time-reversal-invariance non-conserving amplitude, g. In a partial wave
  decomposition the four lowest, parity conserving, transition amplitudes in
  which time-reversal-invariance violation may occur are $^3S_1$ 
  $\leftrightarrow$ $^3D_1$,
  ${^1P_1} \leftrightarrow {^3P_1}$, ${^1D_2} \leftrightarrow {^3D_2}$, ${^3P_2} 
  \leftrightarrow {^3F_2}$. Of these only the latter
  one is allowed for identical nucleons because of the Pauli exclusion 
principle.
  Thus time-reversal-invariance violating effects, if these exist, are strongly
  suppressed in the proton-proton and neutron-neutron systems. As shown by
  Arash, Moravcsik, and Goldstein [9] null tests of time-reversal-invariance
  do not exist in a two-particle in and two-particle out reaction. Observables
  are bilinear product combinations of scattering amplitudes. Tests of
  time-reversal-invariance can be accomplished only through a comparison of
  two distinct observables. As prime example, polarization - analyzing power
  difference determinations, $P - A = -2 \times Im(c^*h + d^*g)/\sigma_0$, are based
  on two independent measurements with their own systematic errors in
  polarimeter analyzing power and beam polarization calibrations. The best
  calibration standards todate carry uncertainties of a few parts in $10^3$.
  Writing the polarization - analyzing power difference as $\epsilon{(1 - D)}/2$,
  with $\epsilon$ the normalized spin-flip differential cross sections difference
  and with D the depolarization parameter, one can show that most such
  difference determinations for the neutron-proton system were made in an
  angular region where D is close to one [10], causing a measure of
  insensitivity to time-reversal-invariance non-conservation. Similarly,
  time-reversal-invariance tests performed in the proton-proton system are
  rather inconclusive.[8] As noted by
  Conzett [11] transmission experiments are not included in the nonexistence
  proof of Arash, Moravcsik, and Goldstein. Indeed, since the total cross
  section can be expressed in terms of the imaginary part of the forward
  scattering amplitude, null tests of time-reversal-invariance can be devised
  for transmission experiments (a selected spin-dependent total cross section
  linearly dependent on the time-reversal-non-invariant amplitude).
  Transmission experiments that test a T-odd / P-even interaction require
  spin 1/2 particles incident on target particles with spin $J \ge 1$ or
  vice versa.
 
    Charge symmetry breaking (CSB) in neutron-proton elastic scattering
  manifests itself as a non-zero difference of the neutron $(A_n)$ and proton
  $(A_p)$ analyzing powers,
  $\Delta{A} = A_n - A_p =  2 \times [Re(b^{*}f) + Im(c^{*}h)]/\sigma_0$.
  The three precision experiments performed (at TRIUMF at 477 MeV [12] and
  at 369 MeV [13], and at IUCF at 183 MeV [14]) have unquestionably shown that
  charge symmetry is broken and that the results for $\Delta{A}$ at the
  zero-crossing angle of the average analyzing power, are very well reproduced
  by meson exchange model calculations (see Fig.1). As shown above
  a T-odd / P-even interaction corresponds to a term in the scattering
  amplitude which is simultaneously charge symmetry breaking.
  Thus, Simonius [15] deduced an upper limit on a T-odd / P-even CSB interaction
  from a comparison of the experimental results with the theoretical predictions
  for the above mentioned three CSB experiments. The upper limit so derived is
  $|\overline{g}_\rho| < 6.7 \times 10^{-3}$ [95\% C.L.]. This is therefore comparable to the
  upper limit deduced from the electric dipole moment of the neutron, taking the
  present experimental limit of $f_\pi$, and is considerably lower than the limits
  inferred from direct tests of a T-odd / P-even interaction. For instance
  the detailed balance experiments give a limit on $|\overline{g}_\rho| < 2.5 
  \times 10^{-1}$.
  [16] As remarked above, it is inconceivable in the Standard Model to account
  for a T-odd / P-even interaction. Nevertheless, there is a need to clarify the
  experimental constraint on a T-odd / P-even interaction by providing
  a BETTER experimental limit. 
 
    Such a better experimental constraint may be provided by an improved upper
  limit on the electric dipole moment of the neutron. In fact a new measurement
  with a sensitivity of $4 \times 10^{-28}$ e.cm has been proposed at the Los Alamos
  Neutron Science Center.[17] But performing an improved n-p elastic scattering
  CSB experiment appears to be a very attractive alternative. One can
  calculate with a great deal of confidence the contributions to CSB due to
  one-photon exchange (the neutron magnetic moment interacting with the current
  of the proton) and due to the n-p mass difference affecting charged one-pion
  and rho-meson exchange. Furthermore, one can select an energy where the
  $\rho^{0}-\omega$ meson mixing contribution changes sign at the same angle where
  the average of the analyzing powers A$_n$ and A$_p$ changes sign and therefore
  does not contribute. This occurs at an incident neutron energy of 320 MeV and
  is caused by the particular interplay of the n-p phase shifts and the form of
  the spin/isospin operator for the $\rho^{0}-\omega$ mixing term. But also
  the contribution due to one-photon excange changes sign at about the same
  angle at 320 MeV. The contribution due to two-pion exchange with an
  intermediate $\Delta$ is expected to be no more than one tenth of the overall
  CSB effect, essentially presenting an upper limit on the theoretical
  uncertainty (see Fig. 2).[18] It has been shown that simultaneous $\gamma-\pi$
  exchanges can only contribute through second order processes and can
  therefore be neglected.[19] Also the effects of inelasticity are neglegibly
  small at 320 MeV. It appears therefore well within reach to reduce the
  theoretical uncertainty in the comparison between experiment and theory.
  Subtracting the calculated difference in the neutron and proton analyzing
  powers from the measured difference in these permits establishing
  an upper limit on a P-even / T-odd / CSB interaction. 
 
    The second TRIUMF experiment measuring CSB in n-p elastic scattering at 347
  MeV obtained the result
  $\Delta$A$ = A_n - A_p =
   ( 59 \pm 7(\rm stat) \pm 7(\rm syst) \pm 2(\rm syst) ) \times 10^{-4}$
  at the zero-crossing angle of the average of $A_n$ and $A_p$. In the experiment
  polarized neutrons were scattered from unpolarized protons and vice versa.
  The polarized (or unpolarized) neutron beam was obtained using the (p,n)
  reaction with a 369 MeV polarized (or unpolarized) proton beam incident on a
  0.20 m long LD$_2$ target. At the TRIUMF energies one makes use
  of the large sideways-to-sideways polarization transfer 
  coefficient $r_t$ at
  9 degrees in the lab. The only difference in obtaining the unpolarized and
  polarized neutron beams was the turning off of the pumping laser light in the
  optically pumped polarized ion source (OPPIS). The polarized proton target
  was of the frozen spin type with butanol beads as target material. The same
  target after depolarization was used as the unpolarized proton target. Great
  care was taken that the two interleaved phases of the experiment were
  performed with identical beam and target parameters except for the
  polarization states. Scattered neutrons and recoiling protons were detected
  in coincidence in the c.m. angular range 53.4 to 86.9 degrees in two
  left-right symmetric detector systems. Rather than measuring $A_n$ 
  and $A_p$
  directly (which would be troubled by not having polarization calibration
  standards of the required precision), the zero-crossings of $A_n$ 
  and $A_p$ were
  determined by fitting the partial angular distributions with polynomials,
  deduced from n-p phase shift analyses. The difference $A_n - A_p$ followed
  by multiplying the difference in the zero-crossing angles by the average slope
  of the analyzing powers (the experiment measured the slope of $A_p$ at the
  zero-crossing angle, which is a very good approximation for the average slope
  at the zero-crossing angle and introduces a negligible error). The principle
  of the measurement is shown schematically in Fig. 3. The execution
  of the experiment depended on a great deal of simultaneous monitoring and
  control measurements (see Fig. 4 for a schematic view of the experiment).  
 
    Both the statistical and systematic errors obtained in the experiment can
  be considerably improved upon. With the OPPIS developments that have taken
  place in the intervening years and with the addition of a biased Na-ionizer
  cell, it will be possible to obtain up to 50 $\mu$A of beam with a
  polarization of 80\% incident on the neutron production LD$_2$ target (a factor
  of 50 increase in neutron beam intensity over the previous CSB
  experiment).[20] A 342 MeV proton beam incident on a 0.20 m long 
  LD$_2$ target
  would present a heat load of 500 W. LH$_2$ targets allowing such heat loads
  have been developed for electron scattering experiments. However, reducing
  the LD$_2$ target thickness to 0.05 m would give a better defined neutron energy
  spectrum reducing the uncertainty in the difference in apparent energies of
  the A$_n$ and $A_p$ measurements. Correcting for the apparent energy difference
  contributes significantly to the systematic error budget. The median of the
  intensity distribution of the proton beam incident on the LD$_2$ target will
  again be kept fixed to within 0.05 mm at two places through feedback loops to
  two sets of steering magnets placed upstream in the beam line. This freezes
  position and direction of the proton beam at the LD$_2$ target, necessary to
  meet the secondary neutron beam energy and direction stability requirement.
  The polarized proton target should again be of the frozen spin type; choosing
  a target material with improved ratio of free protons to bound protons and a
  lower magnetic holding field ($<$~0.2 T) would greatly improve on the
  systematic error by reducing the uncertainty in the background subtraction.
  It appears entirely feasible to reduce the statistical and systematic errors
  by a factor three to four. Such an experiment would constitute a measurement
  of CSB in n-p elastic scattering of unprecedented precision, of great value
  on its own, and would be simultaneously provide the best upper limit on a
  T-odd/P-even interaction. 

\section{Null Tests}
 
    As remarked above true null tests of time-reversal-invariance do not exist
  for spin 1/2 particles scattered from spin 1/2 particles. However, null tests
  exist as transmission measurements for spin 1/2 particles interacting with
  aligned nuclear targets.[11] In such tests one measures the total cross
  section asymmetry A$_{y,xz}$ of vector polarized spin 1/2 particles interacting
  with an aligned nuclear target (for instance a tensor polarized spin 1
  deuteron target). Huffman et al [16] have extracted the five-fold correlation
  parameter A$_{y,xz}$ by observing polarized neutron transmission through
  nuclear spin-aligned $^{165}$Ho. This resulted in an upper limit for
  $|\overline{g}_\rho|$ of $5.9 \times 10^{-2}$ even though the measured value of A$_{y,xz}$ was
  $(8.6 \pm 7.7) \times 10^{-6}$. It is to be noted that only the valence proton in
  $^{165}$Ho contributes to A$_{y,xz}$. 

    Storage rings provide a completely different environment, with the
  advantages outweighing the disadvantages possibly, for high precision tests
  of fundamental symmetries (like testing time-reversal-invariance in the GeV
  range), as discussed for instance by S.E.~Vigdor.[21] At COSY a proton-deuteron
  transmission experiment has been proposed to measure the T-odd/P-even
  observable A$_{y,xz}$ using a polarized internal proton beam (polarization 
  P$_y$)
  and an internal polarized deuterium target P$_{xz}$. Tensor polarized deuterium
  atoms are produced in an atomic beam source based on Stern-Gerlach separation
  in permanent sextupole magnets and adiabatic high frequency transitions.
  Adequate luminosities can be obtained using a window-less storage cell
  placed on the axis of the proton beam [22]. The polarized proton beam is obtained
  from an atomic crossed beam polarized ion source. For this test of a
  T-odd / P-even interaction the COSY ring will serve as accelerator, forward
  spectrometer, and detector.[23] Figure 5 presents a schematic view of the
  COSY facility with the EDDA internal target and detector system for this
  T-odd/P-even test. Crucial to the experiment is a current monitor which can
  register with the required precision the decrease in circulating beam
  intensity with time as function of the circulating proton beam spin state.
  Also of great importance is the precise alignment of the proton beam vector
  and deuteron beam tensor polarizations in order to suppress unwanted spin
  correlation coefficients, which could produce a false result. An accuracy
  of $10^{-6}$ in A$_{y,xz}$ has been planned for the experiment, which would give
  a predicted sensitivity to $|g_\rho|$ of $10^{-3}$ and to $|g_A|$ of 
  $2 \times 10^{-3}$
  for center of mass momenta in the range 200 to 400 MeV/c. The predicted
  sensitivity is decreasing for higher momenta. There is a further measure of
  uncertainty due to the relatively large inelasticity (pion production) in the
  momentum range 2-3 GeV/c, for which the COSY experiment is planned.[24]
  The dilution effect of Coulomb multiple scattering decreases significantly
  with increasing proton beam energy. It will be a tour-de-force for the
  experiment to reach a sensitivity comparable to the present n-p CSB
  experiments. 

\section{K and B Decays}

   Other searches for a T-odd / P-even interaction are made in particle decays,
  e.g., in the decay $K^{+} \to \mu^{+} \pi^{0} \nu_{\mu}$. A non-zero value of
  the muon polarization transverse to the decay plane would be an indication
  of time-reversal-invariance non-conservation. Several experiments have been
  performed using both neutral and charged kaons. There is a unique feature
  to the transverse muon polarization in that it does not have contributions
  from the Standard Model at tree level and that higher order effects are of
  order $10^{-6}$. With only one charged particle in the final state, a final
  state interaction, which can mimic a time-reversal-invariance breaking
  effect, is greatly reduced and is estimated to occur at the same level of
  $10^{-6}$. The more recent effort of measuring the time-reversal-invariance
  non-conserving transverse muon polarization is at KEK using a stopped $K^+$
  beam. The experiment reports a result for P$_T$ = -0.0042 $\pm$ 0.0049(stat)
  $\pm$ 0.0009(syst), based on the data taken in 1996 and 1997, which translates
  into a value of Im$\xi$ = -0.0013 $\pm$ 0.0169(stat) $\pm$ 0.0009(syst).[25] The
  quantity $\xi$ is defined as the ratio of the two form factors, $f_{+}(q^2$) and
  $f_{-}(q^2)$, in the K$_{\mu 3}$ decay matrix element; Im$\xi$ must be equal to
  zero for time-reversal-invariance to hold.[26] With the data allready
  in hand and with the approved data taking time, it is anticipated to arrive
  at a statistical error of $\pm 0.0008$ in Im$\xi$. The best previous experimental
  limits were obtained with both neutral and charged kaons at the BNL-AGS.[27]
  A combination of both experimental results provided a limit on the
  imaginary part of the hadron form factors, Im$\xi$ = -0.01 $\pm$ 0.019. A new
  search for the time-reversal-invariance non-conserving transverse muon
  polarization with in-flight decays $K^+ \to \mu^{+} \pi^{0} 
  \nu_{\mu}$ was
  proposed at the BNL-AGS.[28] It was intended to obtain a sensitivity to the
  tranverse muon polarization of $\pm 0.00013$, corresponding to a sensitivity
  to Im$\xi$ of $\pm 0.0007$. The possibility of similar searches for the
  time-reversal-invariance non-conserving transverse $\tau$ polarization in B
  semileptonic decays, $B \to  M \tau \nu_{\tau}$, has been discussed recently
  by Y.~Kuno.[29] It is estimated that the polarization of the $\tau$ leptons
  could reach as high as as 30\% in $B^+ \to D^{0} \tau \nu$ decays. A non-zero
  value of the transverse muon polarization in $K_{\mu 3}$ decay and of the
  $\tau$ lepton in semi-leptonic B decays would constitute clear evidence for
  new physics. 
  
\section{Summary}

   In summary, searches for a T-odd / P-even interaction have sofar resulted
  in only very modest limits on such an interaction. Most promising are the
  continuing efforts to measure the electric dipole moment of the neutron,
  to measure charge-symmetry breaking in neutron-proton elastic scattering
  at around 320 MeV, and to measure the five-fold correlation parameter
  A$_{y,xz}$ in a proton-deuteron transmission experiment,
  as well as searches of transverse lepton polarizations in K and B decays.

%(not sure of Refs. 26 and 22 Typos)
%not sure of spelling of W.M. Mosse or Morse.

\newpage

%\section{Figures}

%\psdraft
\begin{figure}
\begin{center}
\epsfig{figure=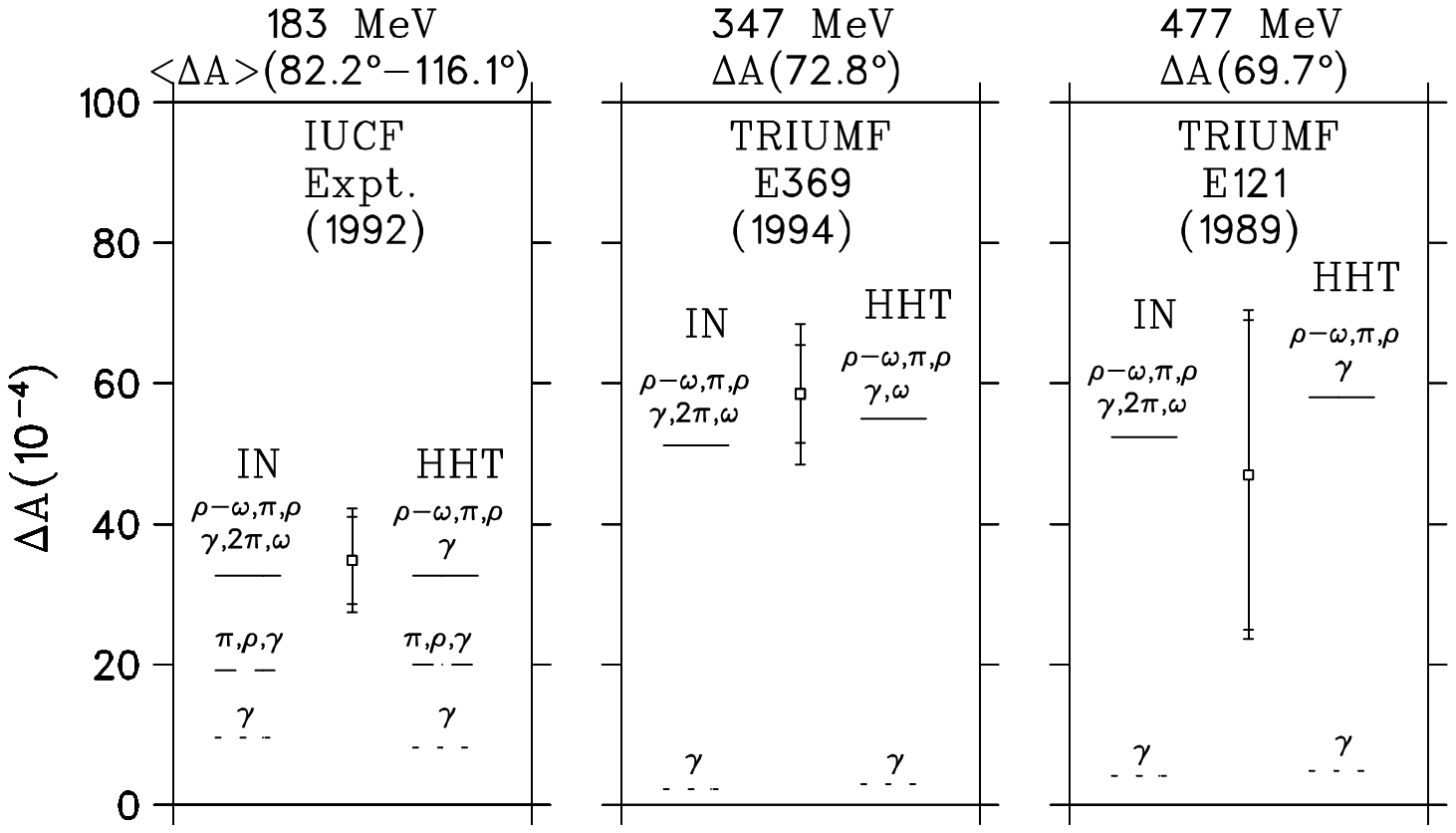,width=\linewidth}
\end{center}
\caption{Experimental results of $\Delta A$ at the zero-crossing angle at
          incident neutron energies of 183, 347, and 477 MeV compared with
          theoretical predictions of Iqbal and Niskanen and Holzenkamp, Holinde,
          and Thomas. The inner error bars present the statistical 
uncertainties;
          the outer error bars have the systematic uncertainties included (added
          in quadrature). For further details see Ref.~13.}
\end{figure}

\begin{figure}
\begin{center}
\epsfig{figure=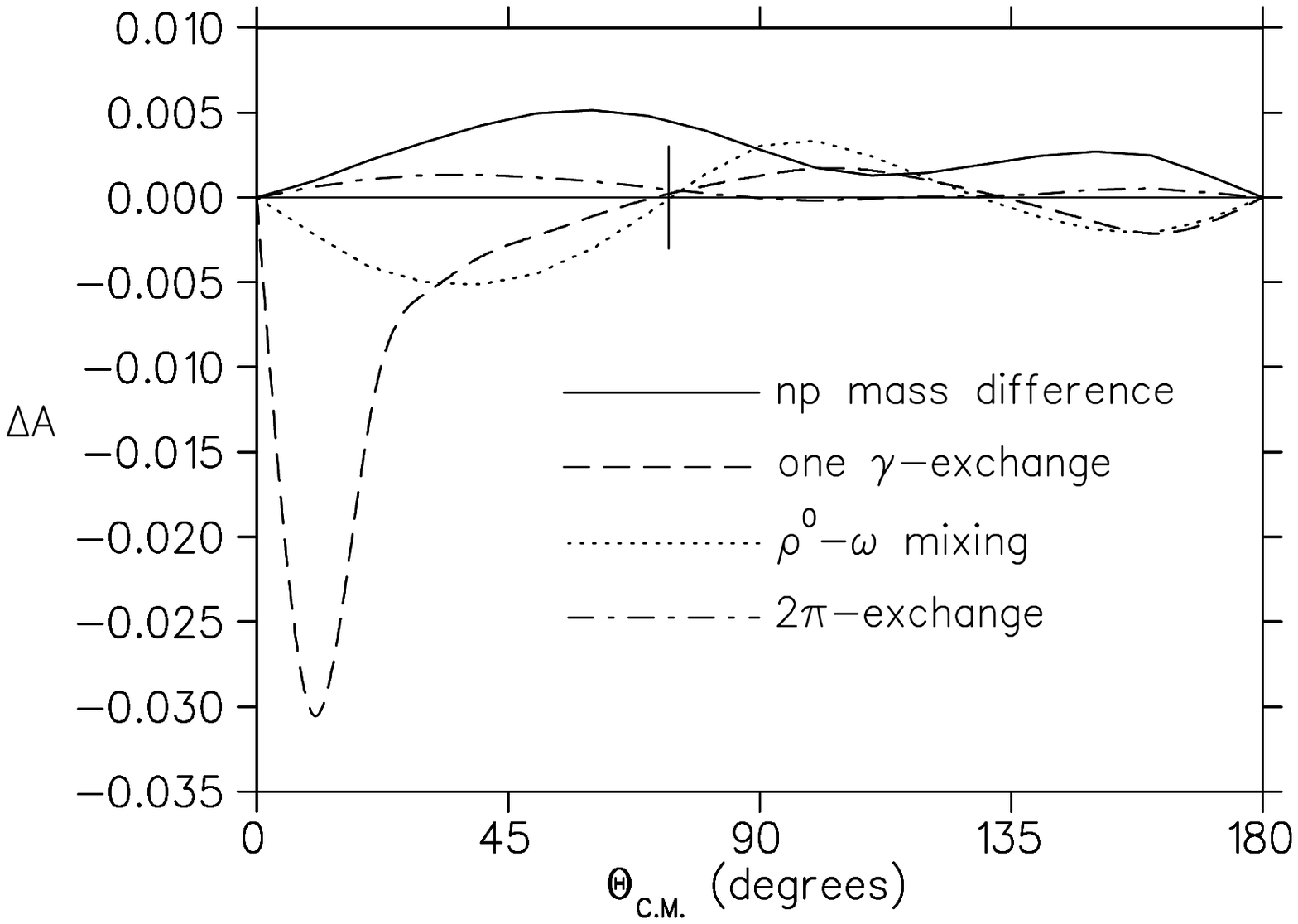,width=0.8\linewidth}
\end{center}
\begin{center}
\epsfig{figure=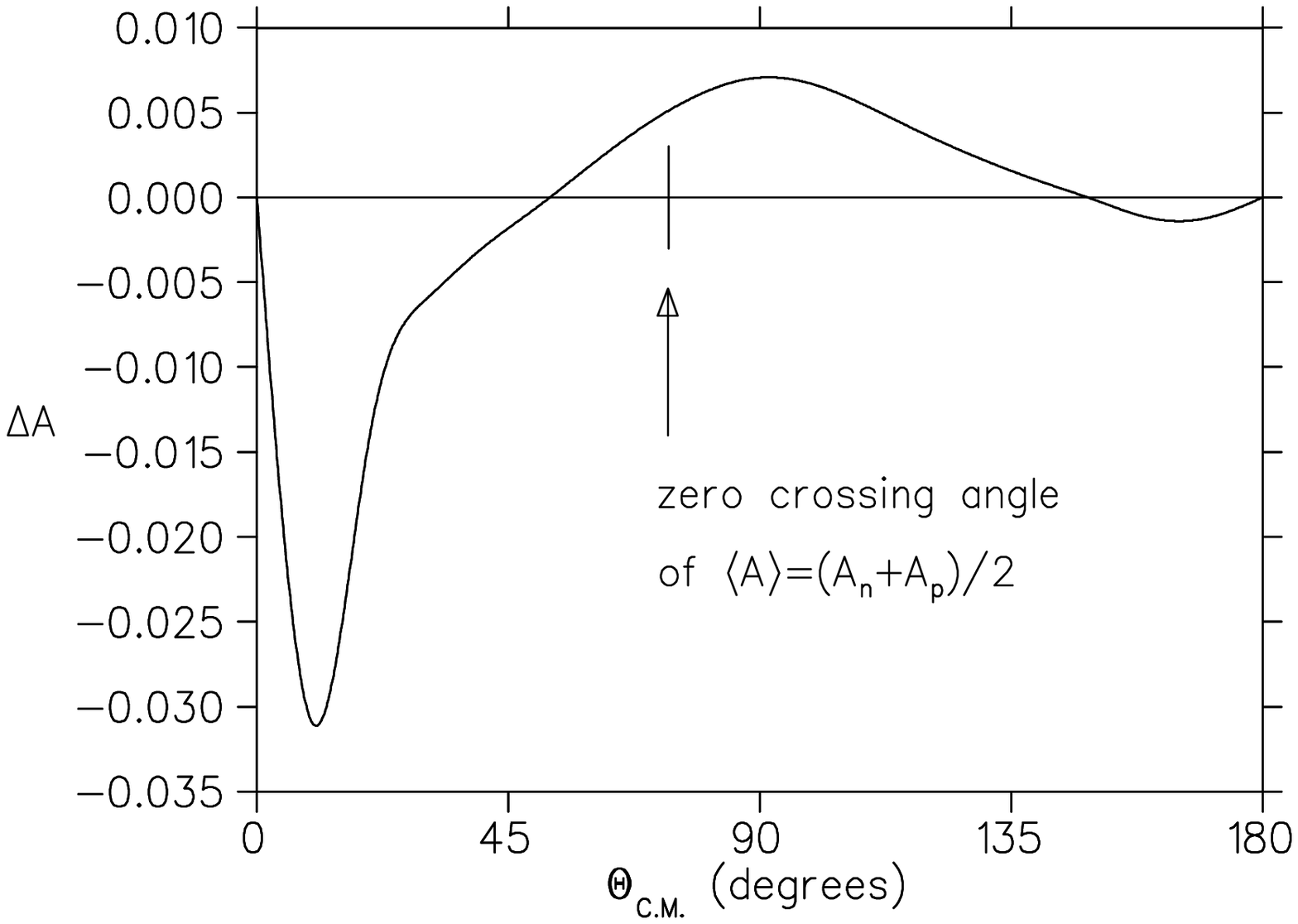,width=0.8\linewidth}
\end{center}
\caption{
    Angular distributions of the different contributions to $\Delta A$ at
          an incident neutron energy of 320 MeV (see Ref.~18). Note that the $\rho^0 - \omega$
          mixing contribution passes through zero at the same angle as the
          average of $A_n$ and A$_p$.}
\end{figure}

\begin{figure}
\begin{center}
\epsfig{figure=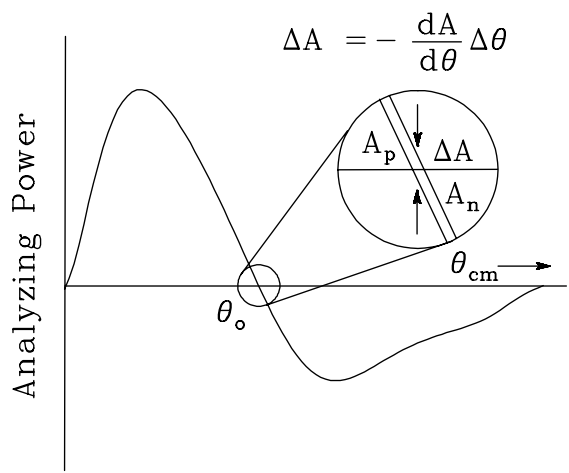,width=0.6\linewidth}
\end{center}
\caption{Principle of the TRIUMF neutron-proton elastic scattering CSB
          experiments.}
\end{figure}

\begin{figure}
\vspace*{-2.0cm}
\begin{center}
\epsfig{figure=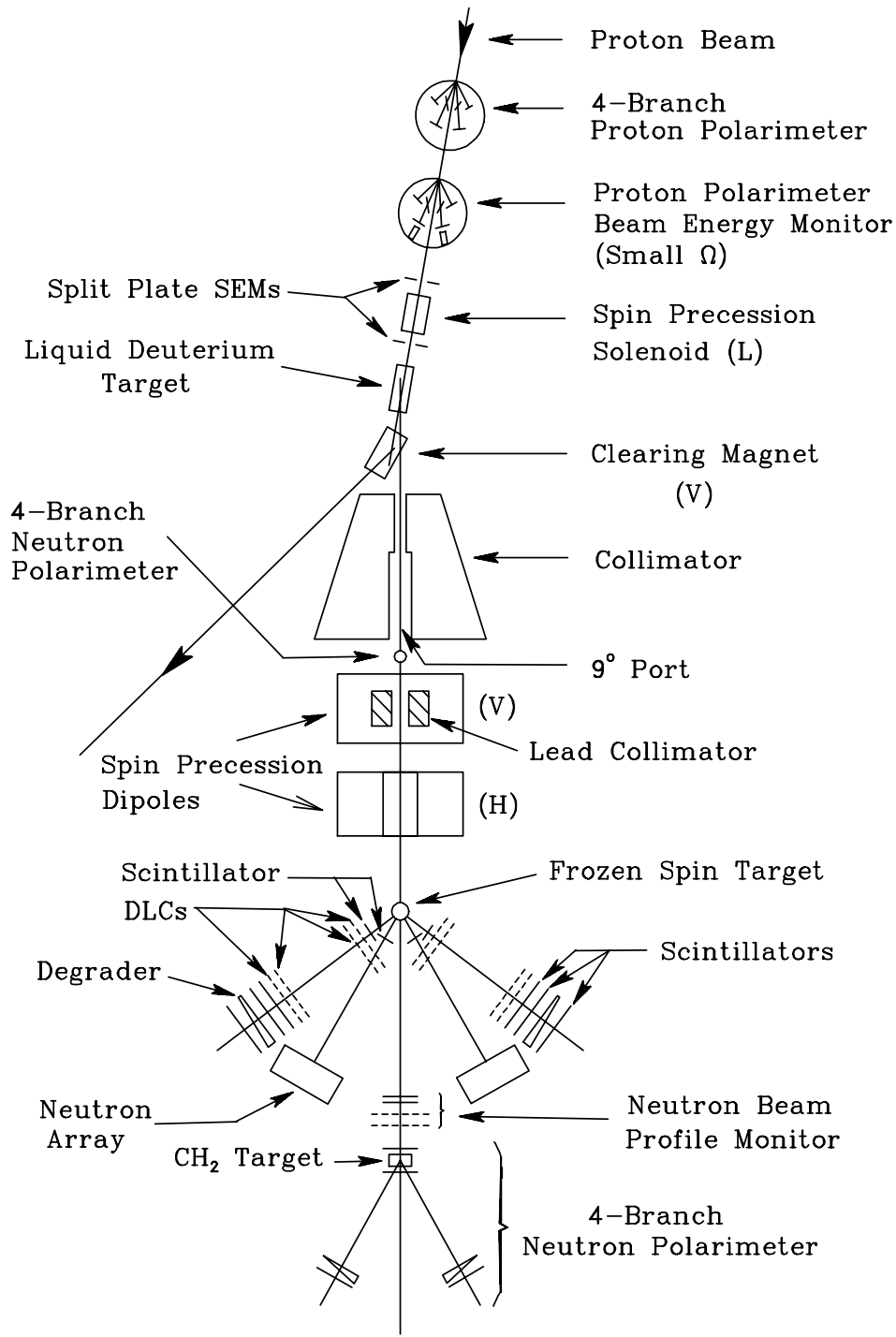,width=\linewidth}
\end{center}
\caption{Schematic view of the TRIUMF CSB experiments.}
\end{figure}
\begin{figure}
\begin{center}
\epsfig{figure=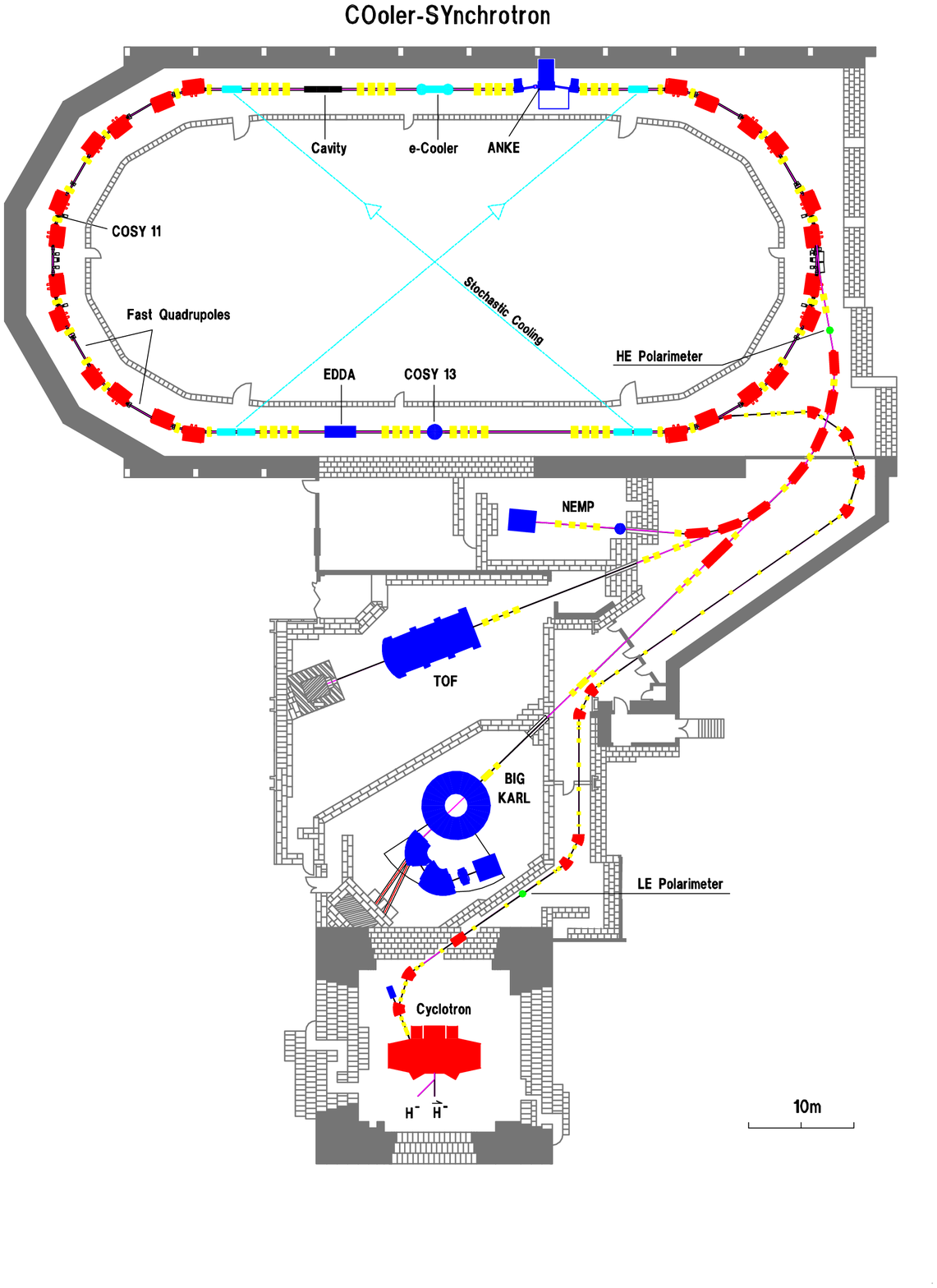,width=1.2\linewidth}
\end{center}
\caption{Schematic view of the COSY facility.}
\end{figure}
 
\end{document}